\pdfoutput=1

\documentclass[11pt,twoside,a4paper,cmspaper,final,collab]{cms-tdr}
\def\svnVersion{168105}\def\cmsCernNo{2012-271}\def\cmsCernDate{\today}\def\cmsMessage{Submitted to the European Physical Journal C}

\begin{document}\cmsNoteHeader{BPH-11-010}

\hyphenation{had-ron-i-za-tion}
\hyphenation{cal-or-i-me-ter}
\hyphenation{de-vices}

\RCS$Revision: 158203 $
\RCS$HeadURL: svn+ssh://svn.cern.ch/reps/tdr2/papers/BPH-11-010/trunk/BPH-11-010.tex $
\RCS$Id: BPH-11-010.tex 158203 2012-11-16 12:50:03Z alverson $
\newlength\cmsFigWidth
\ifthenelse{\boolean{cms@external}}{\setlength\cmsFigWidth{0.95\columnwidth}}{\setlength\cmsFigWidth{0.45\textwidth}}
\ifthenelse{\boolean{cms@external}}{\providecommand{\cmsLeft}{top\xspace}}{\providecommand{\cmsLeft}{left}}
\ifthenelse{\boolean{cms@external}}{\providecommand{\cmsRight}{bottom\xspace}}{\providecommand{\cmsRight}{right}}

\cmsNoteHeader{BPH-11-010} 

\newcommand\usedlumi {4.6\fbinv}
\newcommand{\chic}{\ensuremath{\chi_{\mathrm{c}}}\xspace}
\newcommand{\Chizero}{\ensuremath{\chi_{\mathrm{c}0}}\xspace}
\newcommand{\Chione}{\ensuremath{\chi_{\mathrm{c}1}}\xspace}
\newcommand{\Chitwo}{\ensuremath{\chi_{\mathrm{c}2}}\xspace}
\newcommand{\Chionetwo}{\ensuremath{\chi_{\mathrm{c}1,2}}\xspace}
\newcommand{\theratio}{\ensuremath{N_{\Chitwo}/N_{\Chione}}\xspace}
\newcommand{\eoneetwo}{\ensuremath{{\varepsilon_{1}/\varepsilon_{2}}}\xspace}
\ifthenelse{\boolean{cms@external}}{\providecommand{\cmsBreak}{\linebreak[4]}}{\providecommand{\cmsBreak}{relax}}

\title{Measurement of the relative prompt production rate of $\Chitwo$ and $\Chione$ in pp collisions  at $\sqrt{s} = 7$\TeV}

\date{\today}

\abstract{ A measurement is presented of the relative prompt production rate of \Chitwo and \Chione  with 4.6\fbinv of data collected by the CMS experiment at the LHC in pp collisions at $\sqrt{s}= 7\TeV$. The two states are measured via their radiative decays $ \chic \to \cPJgy + \gamma$, with the photon converting into an $\Pep\Pem$ pair for \cPJgy\ rapidity $|y(\cPJgy)| < 1.0$ and photon transverse momentum $\pt(\gamma) > 0.5\GeVc$. The measurement is given for six intervals of $\pt(\cPJgy)$ between 7 and 25\GeVc. The results are compared to theoretical predictions.
}

\hypersetup{%
pdfauthor={CMS Collaboration},%
pdftitle={Measurement of the relative prompt production rate of chi(c2) and chi(c1) in pp collisions  at sqrt(s) = 7 TeV},%
pdfsubject={CMS},%
pdfkeywords={CMS, physics}}

\maketitle 

\section{Introduction}
Understanding charmonium production in hadronic collisions is a
challenge for quantum chromodynamics (QCD). The \JPsi production cross section
measurements at the Tevatron~\cite{cdf:jpsi,cdf:jpsi2}  were found to
disagree by about a factor of 50 with theoretical color-singlet calculations~\cite{theory:production}.
Soon after, the CDF experiment reported a $\Chitwo/ \Chione$
cross section ratio that extended up to $\pt(\JPsi) \simeq 10\GeVc$, where \pt is the transverse momentum, and favored $\Chione$ production over $\Chitwo$~\cite{cdf:chi2chi1}. The cross section ratio was also studied recently at the Large Hadron Colllider (LHC) in Ref.~\cite{lhcb:chi}. These measurements independently
suggest that charmonium production cannot be explained through relatively
simple models.

This paper presents a measurement of the prompt \cmsBreak $\Chitwo/ \Chione$
cross section ratio by the Compact Muon Solenoid (CMS)
experiment at the LHC in pp collisions at a
center-of-mass energy of 7\TeV. Prompt refers to the production of $\chic$ mesons that originate from the primary pp interaction point, as opposed to the ones from the decay of B hadrons. Prompt  production includes both directly produced $\chic$ and also indirectly produced $\chic$ from the decays of short-lived intermediate states, e.g. the radiative decay of the $\psi(2S)$. The measurement is based on the
reconstruction of the $\chic$ radiative decays to $\JPsi + \gamma$, with the
low transverse momentum  photons (less than 5\GeVc)  being detected through their
conversion into electron-positron pairs. The analysis uses data collected
in 2011, corresponding to a total integrated luminosity of \usedlumi. When estimating acceptance and efficiencies, we assume that the $\Chitwo$ and $\Chione$ are produced unpolarized, and we supply the correction factors needed to modify the results for several different polarization scenarios.

Due to the extended reach in transverse momentum made possible by the LHC energies,
the cross section ratio measurement is expected to discriminate between different
predictions, such as those provided by the \cmsBreak \kt-factorization \cite{baranov} and next-to-leading order nonrelativistic QCD (NRQCD) \cite{nlonrqcd} theoretical approaches.

The strength of the ratio measurement is that most theoretical uncertainties cancel, including the quark  masses, the value of the strong coupling constant $\alpha_s$, as well as experimental uncertainties on quantities such as integrated luminosity, trigger efficiencies, and, in part, reconstruction efficiency. Therefore, this ratio can be regarded as an important reference measurement to test the validity of various theoretical quarkonium production models. With this paper, we hope to provide further guidance for future calculations.

\section{CMS detector}

A detailed description of the CMS apparatus is given in Ref.~\cite{JINST}. Here we  provide a short summary of the detectors relevant for this measurement.

The central feature of the CMS apparatus is a superconducting solenoid of 6\unit{m} internal diameter. Within the field volume are the silicon pixel and strip tracker, the crystal electromagnetic calorimeter  and the brass/scintillator hadron calorimeter. Muons are measured in gas-ionization detectors embedded in the steel return yoke. In addition to the barrel and endcap detectors, CMS has extensive forward calorimetry.

The inner tracker measures charged particles within the pseudorapidity range $|\eta| < 2.5$,  where $\eta =  -\ln[\tan(\theta/2)]$, and  $\theta$ is the polar angle measured from the beam axis. It consists of 1440 silicon pixel and 15\,148 silicon strip detector modules. In the central region, modules are arranged in 13 measurement layers. It provides an impact parameter resolution of ${\sim}15\mum$.

Muons are measured in the pseudorapidity range $|\eta|< 2.4$, with detection planes made using three technologies: drift tubes, cathode strip chambers, and resistive plate chambers. Matching the muons to the tracks measured in the silicon tracker results in a transverse momentum resolution between 1 and 1.5\%, for \pt values up to 50\GeVc.

The first level (L1) of the CMS trigger system, composed of custom hardware processors, uses information from the calorimeters and muon detectors to select the most interesting events. The high-level trigger (HLT) processor farm further decreases the event rate from around 100\unit{kHz} to around 300\unit{Hz}, before data storage. The rate of HLT triggers relevant for this analysis was in the range 5--10\unit{Hz}. We analyzed about 60 million such triggers.

\section{Experimental method} \label{sec:expmethod}
We select $\Chione$ and $\Chitwo$ candidates by searching for their radiative
decays into the $\JPsi + \cPgg$ final state, with the \JPsi decaying
into two muons.
The $\Chizero$ has too small a branching fraction into this final state to perform a useful measurement, but we consider it in the modeling of the signal lineshape.
Given the small difference between the $\JPsi$ mass, $3096.916\pm0.011\MeVcc$, and the $\Chione$ and $\Chitwo$ masses, $3510.66\pm0.07\MeVcc$ and $3556.20\pm0.09\MeVcc$, respectively~\cite{PDG}, the detector must be able to reconstruct photons of low transverse momentum. In addition, excellent photon momentum resolution is needed to resolve the two states. In the center-of-mass frame of the $\chic$ states, the photon has an energy of
390\MeV when emitted by a $\Chione$ and 430\MeV when emitted by a
$\Chitwo$. This results in most of  the photons having a \pt in the laboratory frame smaller than 6\GeVc.
The precision of the cross section ratio measurement depends crucially on the experimental photon energy resolution, which must be good enough to separate the two states. A very accurate measurement of the photon energy is obtained by measuring electron-positron pairs originating from a photon conversion in the beampipe or the inner layers of the silicon tracker. The superior resolution of this approach, compared to a calorimetric energy measurement, comes at the cost of a reduced yield due to the small probability for a conversion to occur in the
innermost part of the tracker detector and, more importantly, by the small
reconstruction efficiency for low transverse momentum tracks whose origin is displaced  with respect to the beam axis. Nevertheless, because of the high $\chic$ production cross section at the LHC, the use of conversions leads to the most precise result.

For each $\chi_{\mathrm{c}1,2}$ candidate, we evaluate the mass difference
$\Delta m = m_{\mu\mu \gamma} - m_{\mu \mu}$ between the dimuon-plus-photon invariant mass, $m_{\mu\mu \gamma}$, and the dimuon invariant mass, $m_{\mu \mu}$. We use the quantity $Q = \Delta m + m_{\JPsi}$, where $m_{\JPsi}$ is the world-average mass of the \JPsi from Ref.~\cite{PDG}, as a convenient variable for plotting the invariant-mass distribution. We perform an unbinned maximum-likelihood fit to the $Q$ spectrum to extract the yield of prompt $\Chione$ and $\Chitwo$ as
a function of the transverse momentum of the \JPsi. A correction is applied
for the differing acceptances for the two states. Our results are given in terms of the prompt production ratio $R_\mathrm{p}$, defined as

\[
R_\mathrm{p} \equiv
\frac{\sigma(\Pp\Pp \to \Chitwo +X )  \mathcal{B}(\Chitwo \to \JPsi + \gamma) }{ \sigma(\Pp\Pp \to \Chione +X )  \mathcal{B}(\Chione \to \JPsi + \gamma) }
=\frac{N_{\Chitwo}}{N_{\Chione}} \cdot\frac {\varepsilon_1}{\varepsilon_2} ,
\]
where $\sigma(\Pp\Pp \to \chic +X )$  are the $\chic$ production cross sections, $\mathcal{B}(\chic \to \JPsi + \gamma)$ are the $\chic$ branching fractions, $N_{\chi_i}$ are the number of candidates of each type obtained
from the fit, and  $\eoneetwo$ is the ratio of the efficiencies for the two $\chic$  states. The branching fractions
$\mathcal{B}(\chi_{\mathrm{c}1,2} \to \JPsi + \gamma)$, taken from Ref.~\cite{PDG}, are also used to calculate the ratio of production cross sections.

\section{Event reconstruction and selection}

In order to select $\chic$ signal events, a dimuon trigger is used to record events containing the decay $\JPsi \to \mu \mu$. The L1 selection requires two muons without an explicit constraint on their transverse momentum. At the HLT, opposite-charge dimuons are reconstructed and the dimuon rapidity $y(\mu \mu)$ is required to satisfy $|y(\mu \mu)|<1.0$, while the dimuon \pt must exceed a threshold that increased from 6.5 to 10\GeVc as the trigger configuration evolved to cope with the instantaneous luminosity increase. Events containing dimuon candidates with invariant mass from 2.95 to 3.25\GeVcc are recorded. Our data sample consists of events where multiple pp interactions occur. At each bunch crossing, an average of six primary vertices is reconstructed, one of them related to the interaction that produces the $\chic$ in the final state, the others related to softer collisions (pileup).

In the \JPsi selection, the muon tracks are required to pass the following criteria. They must have at least 11 hits in the tracker, with at least two in the pixel layers, to remove background from decays-in-flight. The $\chi^2$ per degree of freedom of the track fit must be less than 1.8. To remove background from cosmic-ray muons, the tracks must intersect a cylindrical volume  of radius 4\cm and total length 70\cm, centered at the nominal interaction point and  with its axis parallel to the beam line. Muon candidate tracks are required to have $\pt> 3.3$\GeVc, $|\eta| \leq 1.3$ and match a well-reconstructed segment in at least one muon detector~\cite{muoncraft}.
Muons with opposite charges are paired. The two muon trajectories are fitted with a common vertex constraint, and events are retained if the fit $\chi^2$ probability is larger than 1\%. If more than one muon pair is found in an event, only the pair with the largest vertex $\chi^2$ probability is selected.
For the final $\Chione$ and $\Chitwo$ selection, a dimuon candidate must have an invariant mass  between 3.0 and 3.2\GeVcc and  $|y|<1.0$.

In order to restrict the measurement to the prompt \JPsi signal component,
the pseudo-proper decay length of the \JPsi ($\ell_{\JPsi}$), defined as $\ell_{\JPsi} = L_{xy}\cdot m_{\JPsi}/\pt(\JPsi)$, where $L_{xy}$ is the most probable transverse decay length in the laboratory frame  ~\cite{cms:jpsi}, is required to be less than 30~$\mu$m. In the region $\ell_{\JPsi} < 30\mum$, we estimate, from the observed $\ell_{\JPsi}$ distribution, a contamination of the nonprompt component (originating from the decays of \PB\ hadrons) of about 0.7\%, which has a negligible impact on the total systematic uncertainty.

To reconstruct the photon from radiative decays,  we use the tracker-based conversion reconstruction described in Refs.~\cite{TRK-10-001,trk10001,TRK-10-003}. We
summarize the method here, mentioning the further requirements needed to
specialize the  conversion reconstruction algorithm to the $\chi_c$
case. The algorithm relies on the capability of iterative tracking to efficiently reconstruct displaced and low transverse momentum tracks. Photon conversions are characterized by an electron-positron pair originating from a common vertex. The $e^+ e^-$ invariant mass must be consistent with zero within its uncertainties and the two tracks are required to be parallel at the conversion point.

Opposite-sign track pairs are first required to have more than four hits and a normalized
$\chi^2$ less than 10. Then
the reconstruction algorithm  exploits the conversion-pair
signature to distinguish between genuine and misidentified background pairs. Information from the calorimeters is not used for conversion reconstruction in our analysis.
The primary pp collision vertex associated with the photon conversion, see below, is required to lie outside both track helices. Helices projected onto the transverse plane form circles; we define $d_\text{m}$ as the distance between the  centers of the two circles minus the sum of their radii. The value of $d_\text{m}$ is negative when the two projected trajectories intersect. We require the condition $-0.25 < d_\text{m} < 1.0 \cm$ to be satisfied. From simulation, we have found that most of the electron-positron candidate pair background comes from misreconstructed track pairs originating from the primary vertex. These typically have negative  $d_\text{m}$ values, thus explaining the asymmetric $d_\text{m}$ requirements.

In order to reduce the contribution of misidentified
conversions from low-momentum displaced tracks that are artificially propagated back to the silicon tracker, the two candidate conversion tracks must have one of their two innermost hits in the same silicon tracker layer.

The distance along the beam line between the extrapolation of each conversion track candidate and the nearest reconstructed event vertex must be less than five times its estimated uncertainty. Moreover, among the two event vertices closest to each track along the beam line, at least one vertex must be in common

A reconstructed primary vertex is assigned to the reconstructed
conversion by projecting the photon momentum onto the beamline and choosing the
closest vertex along the beam direction. If the value of the distance is larger than five times its estimated uncertainty, the photon candidate is rejected.

The primary vertex associated with the conversion is required to be
compatible with the reconstructed $\JPsi$ vertex. This requirement is
fulfilled when the three-dimensional distance between the two
vertices is compatible with zero within five standard
deviations. Furthermore, a check is made that neither of
the two muon tracks used to define the $\JPsi$ vertex  is used as one of the  conversion track pair.

The $\Pep\Pem$ track pairs surviving the selection are then fitted to a common vertex with a
kinematic vertex fitter that constrains the
tracks to be parallel at the vertex in both the transverse and longitudinal planes.
The pair is retained if the fit $\chi^2$ probability
is greater than 0.05\%.
If a track is shared among two or
more reconstructed conversions, only the conversion with
the larger vertex $\chi^2$ probability is retained.

Only reconstructed conversions with
transverse distance of the vertex from the center of the mean pp collision position larger than $1.5\cm$
are considered. This requirement suppresses backgrounds
caused by track pairs originating from the primary event vertex that might
mimic a conversion, such as from $\Pgpz$ Dalitz decay, while
retaining photon conversions  occurring within the beampipe.

Finally, each conversion candidate is associated with every other
conversion candidate in the event, and with any photon reconstructed using calorimeter information.
Any pairs of conversions or conversion plus photon with an invariant
mass between 0.11 and 0.15\GeVcc, corresponding to a two-standard-deviation  window around the  $\Pgpz$ mass, is rejected.
We have verified that the $\Pgpz$ rejection requirement, while effectively reducing the
background, does not affect the $R_\mathrm{p}$ measurement within its uncertainties.

Converted photon candidates are required to have $\pt > 0.5 $\GeVc , while no requirement is imposed on the pseudorapidity  of the photon.

\begin{figure}[!hbtp]
\centering
\includegraphics[width=\cmsFigWidth]{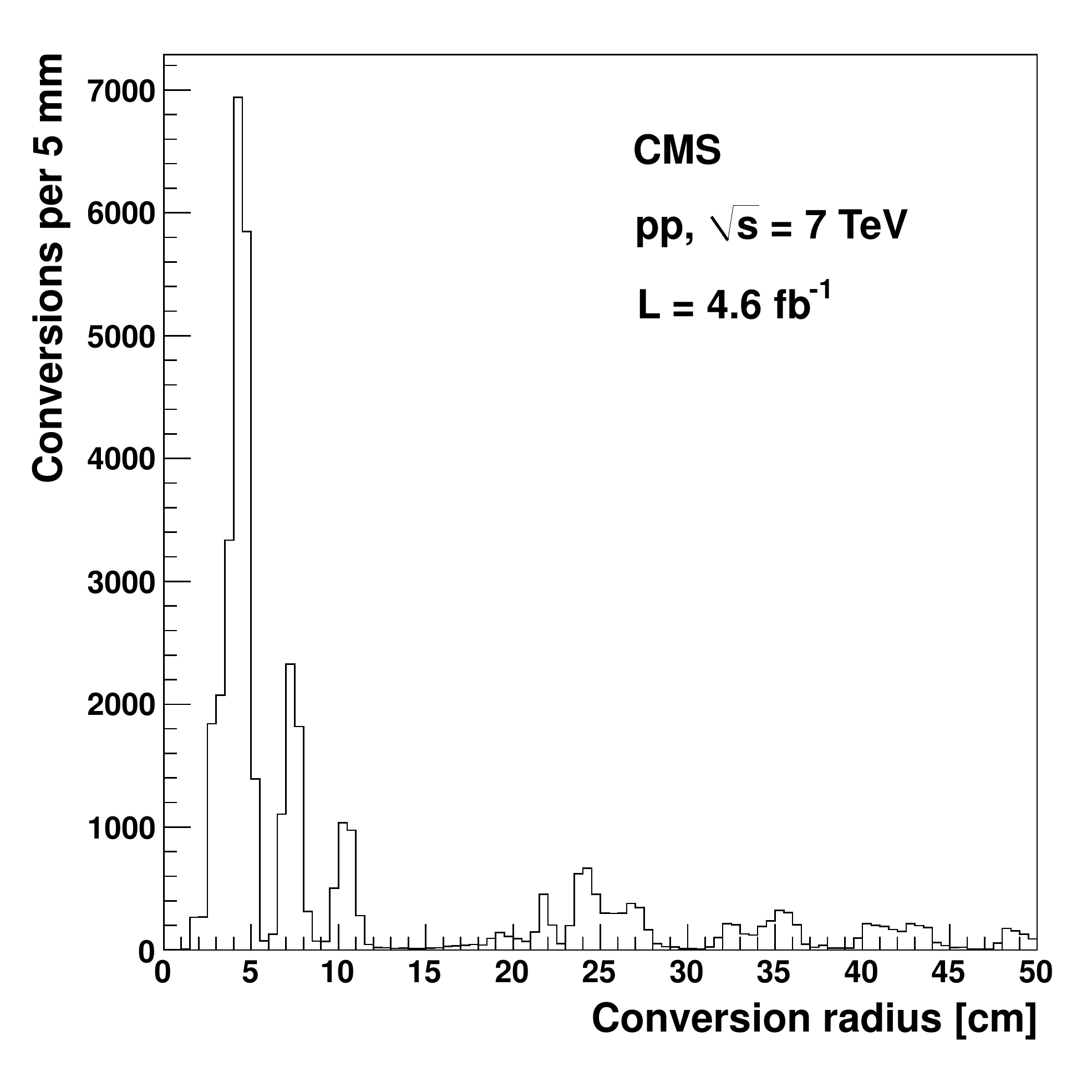}
\caption{Distribution of the conversion radius for the $\chic$ photon candidates.}
\label{fig:convradius}
\end{figure}

The distribution of the photon conversion radius for $\chic$ candidates is
shown in Fig.~\ref{fig:convradius}. The first peak corresponds to the beampipe and first pixel barrel layer, the second and third peaks correspond to the two outermost pixel layers, while the remaining features at radii larger than $20\cm$ are due to the four innermost silicon strip layers. The observed distribution of the photon conversion radius is consistent with the known distribution of material in the tracking volume and with Monte Carlo simulations ~\cite{TRK-10-003}.

\section{Acceptance and efficiencies}
\label{sec:efficiency}

In the evaluation of $R_\mathrm{p}$, we must take into account the possibility
that the geometric acceptance  and the  photon reconstruction efficiencies are
not the same for $\Chione$ and $\Chitwo$.

In order to determine the acceptance correction, a \ifthenelse{\boolean{cms@external}}{\linebreak[4]}{}Monte Carlo (MC)
simulation  sample of equal numbers of $\Chione$ and $\Chitwo$ has been
used. This sample was produced using a  \PYTHIA~\cite{pythiacit} single-particle  simulation in which a  $\Chione$ or $\Chitwo$ is
generated with a transverse momentum  distribution produced from a parameterized fit to the CMS measured $\psi(2S)$ spectrum~\cite{cms:jpsi2}. The use of the $\psi(2S)$ spectrum is motivated by the proximity of the  $\psi(2S)$ mass to the states under examination. The impact of this choice is discussed in Section \ref{sec:systematics}.

Both $\chic$ states in the simulation are forced to decay to $\JPsi + \cPgg$ isotropically in their rest frame, i.e., assuming they are produced unpolarized. We discuss later the impact of this assumption. The  decay products are then processed through the full CMS detector
simulation, based on \GEANTfour~\cite{Geant4_1, Geant4_2}, and subjected to the trigger emulation and the full event reconstruction. In order to produce the most realistic  sample of simulated $\chic$ decays, digitized signals from MC-simulated inelastic pp  events are mixed with those from simulated  signal tracks. The number of inelastic events to mix with each signal event is sampled from a Poisson distribution to accurately reproduce the amount of  pileup in the data.

The efficiency ratio $\eoneetwo$ for different \JPsi transverse momentum bins is determined using :
\[
\frac{\varepsilon_1}{\varepsilon_2}  =
\frac{N_{\Chione}^\text{rec}}{N_{\Chione}^\text{gen}} /  \frac{N_{\Chitwo}^\text{rec}}{N_{\Chitwo}^\text{ gen}},
\]

where $N^\text{gen}$ is the number of $\chic$ candidates generated in the MC simulation within the kinematic range $|y(\JPsi)| < 1.0$, $\pt(\gamma)>0.5\GeVc$,
and $N^\text{rec}$ is the number of candidates reconstructed with the
selection above. The resulting values are shown in Table~\ref{table:eeonetwo}, where the uncertainties are statistical only and determined from the MC sample assuming binomial distributions. The increasing trend of $\eoneetwo$ is expected, because $\pt(\JPsi)$ is correlated with the \pt of the photon, and at higher photon \pt our conversion  reconstruction efficiency is approximately constant. Therefore, efficiencies  for the $\Chione$ and the $\Chitwo$  are approximately the same at high $\pt(\JPsi)$.

This technique also provides an estimate of the absolute $\chic$ reconstruction efficiency, which is given by the product of the photon conversion probability, the  $\chic$ selection efficiency, and, most importantly, the conversion reconstruction efficiency, which corresponds to the dominant contribution. This product varies  as a function of $\pt(\gamma)$, and goes from $4 \times  10^{-4}$ at 0.5\GeVc to around $10^{-2}$ at 4\GeVc, where it saturates.

\begin{table}[htbp]
     \centering
     \topcaption{Ratio of  efficiencies $\eoneetwo$ as a function of the \JPsi transverse momentum from MC simulation. The uncertainties are statistical only.}
     \label{table:eeonetwo}
     \begin{tabular}{cc}
        \hline
        $\pt(\JPsi) [\GeVcns]$& $\eoneetwo$  \\

        \hline
        7--9 & 0.903   $\pm $ 0.023   \\
        9--11 & 0.935  $\pm $ 0.019  \\
        11--13 & 0.945 $\pm $ 0.021 \\
        13--16 & 0.917 $\pm $ 0.022 \\
        16--20 & 0.981 $\pm $ 0.031 \\
        20--25 & 1.028 $\pm $ 0.049 \\
        \hline

     \end{tabular}

\end{table}

\section{Signal extraction}

We extract the numbers of $\Chione$ and $\Chitwo$ events, $N_{\Chione}$ and
$N_{\Chitwo}$, respectively, from the data by performing an unbinned maximum-likelihood fit to the Q spectrum in various ranges of \JPsi transverse momentum.

Because of the small intrinsic width of the $\chic$ states
we are investigating, the observed signal shape is dominated by the
experimental resolution.
The signal probability density function (PDF) is derived from the MC simulation described in Section \ref{sec:efficiency}, and is modeled by  the superposition of two  double-sided Crystal Ball
functions  \cite{xball} for  the $\Chione$ and
$\Chitwo$ and a single-sided Crystal Ball function for the $\Chizero$.
Each double-sided Crystal Ball function consists of a Gaussian core with exponential tails on both the high- and low-mass sides. We find this shape to provide an accurate parameterization of the $Q$  spectra derived from MC simulation. When fitting the data, we fix all the parameters of the Crystal Ball function to the values that best fit our MC simulation and use a  maximum-likelihood approach to derive $N_{\Chione}$ and $N_{\Chitwo}$, which are the integrals of the PDFs for the two resonances. Because the Q resolution depends on the \pt of the \JPsi, a set of shape parameters is determined for each bin of $\pt(\JPsi)$. Simulation shows that the most important feature of the $\Chizero$ signal shape is the low-mass tail due to radiation from the electrons, while the high-mass tail is overwhelmed by the combinatorial background and the low-mass tail of the other resonances. Hence  the choice to use a single-sided Crystal Ball function to fit the $\Chizero$ mass distribution. Different choices  of the $\Chizero$ signal parameterization are found to cause variations in the measured $R_\text{p}$ values that are well within the quoted systematic uncertainties given below.

The background is modeled by a probability distribution function defined as
\[
  N_{bkg}(Q)= (Q-q_0)^{\alpha_1} \cdot e^{(Q-q_0) \cdot \beta_1},
\]
where $\alpha_1$ and $\beta_1$ are free parameters in the fit, and $q_0$ is set to 3.2\GeVcc.

\begin{figure}[thb]
    \centering
    \includegraphics[width=\cmsFigWidth]{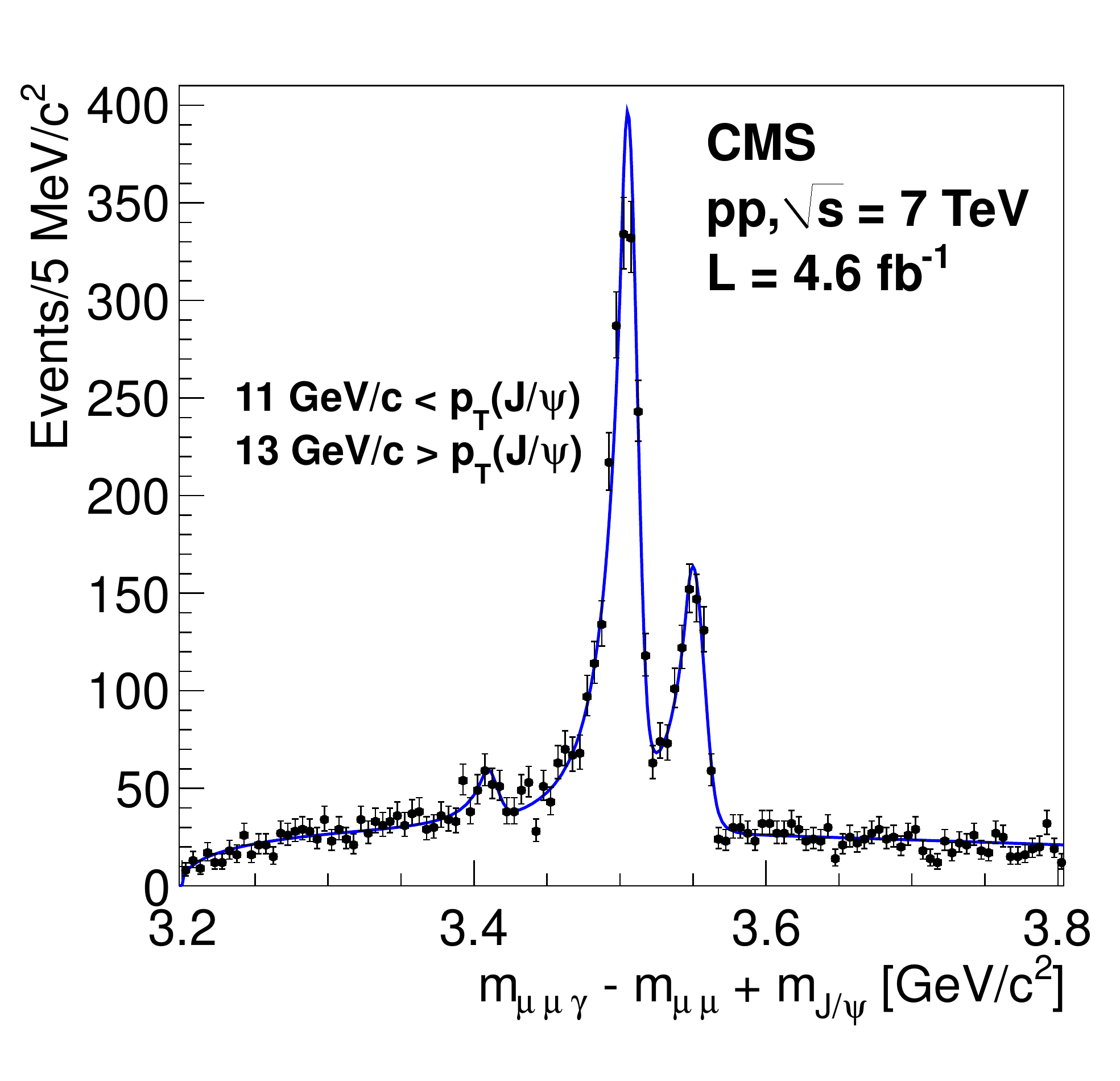}
    \includegraphics[width=\cmsFigWidth]{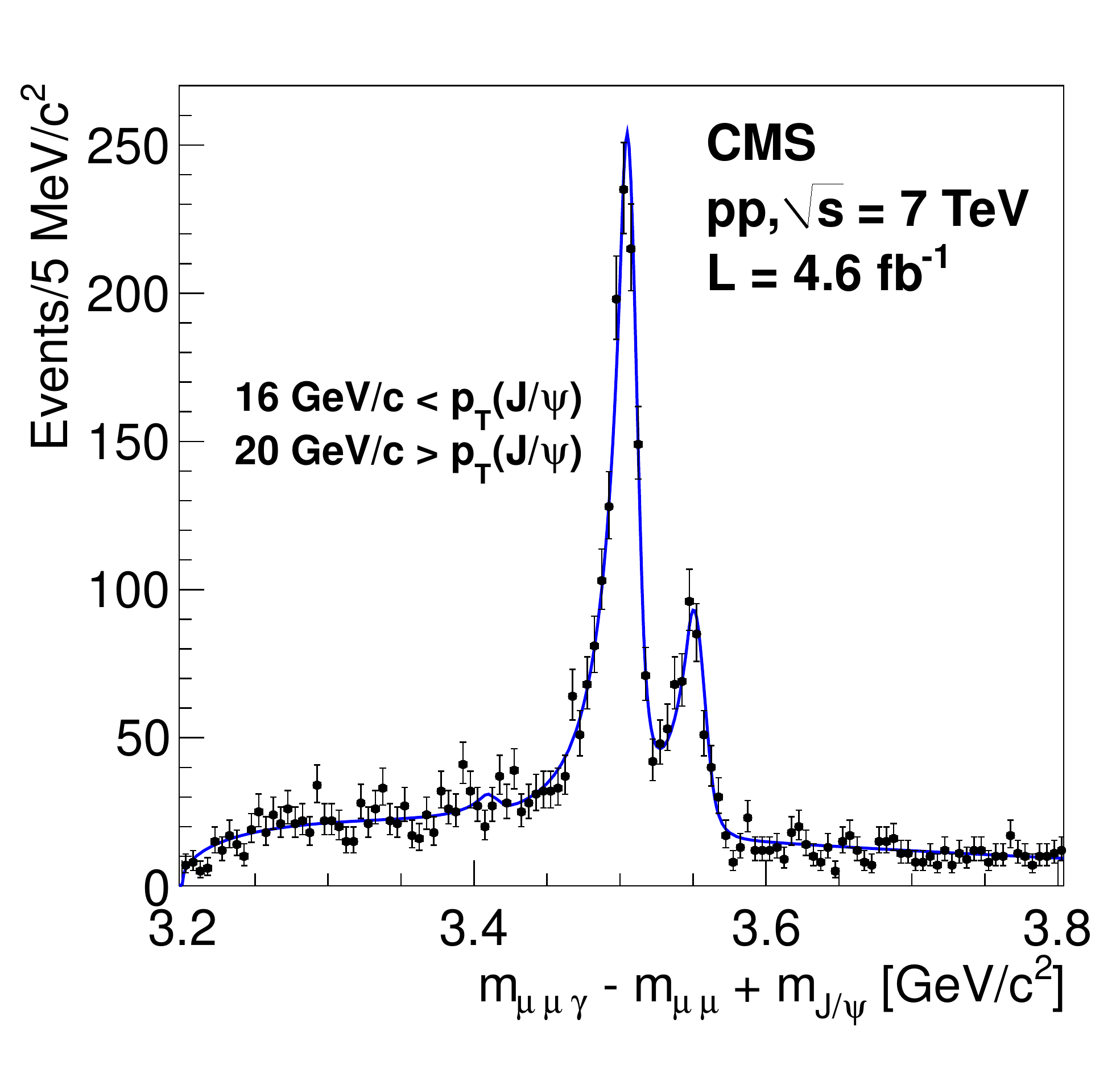}
    \caption{The distribution of the variable $Q = m_{\mu\mu \gamma} - m_{\mu \mu}  + m_{\JPsi}$  for $\chi_c$ candidates with $\pt(\JPsi)$ ranges shown in the figures. The line shows the fit to the data.}
    \label{fig:masspectra}
\end{figure}

In Fig.~\ref{fig:masspectra} we show the Q distribution for two different ranges, $11 <\pt(\JPsi)<  13\GeVc$ (\cmsLeft) and $16 <\pt(\JPsi)<20\GeVc$ (\cmsRight).
This procedure is repeated for several ranges in the transverse momentum of the \JPsi in
order to extract $N_{\Chione}$ and $N_{\Chitwo}$ in the corresponding bin.

The results are shown in Table~\ref{table:rawn}, where the reported
uncertainties are statistical only.

\begin{table}[htbp]
\begin{center}
\topcaption{Numbers of $\Chione$ and $\Chitwo$ events extracted from the maximum-likelihood fit, and the ratio of the two values. Uncertainties are statistical only.}
 \label{table:rawn}
\begin{tabular}{cccc}
\hline
 $\pt(\JPsi)$ [\GeVcns]& $N_{\Chione}$  & $N_{\Chitwo}$  & $N_{\Chitwo} / N_{\Chione}$\\
\hline

7--9 & 618 $\pm$ 31 & 315 $\pm$ 24 & 0.510 $\pm$ 0.049     \\
9--11 & 1680 $\pm$ 49 & 788 $\pm$ 37 & 0.469 $\pm$ 0.027   \\
11--13 & 1819 $\pm$ 51 & 819 $\pm$ 38 & 0.451 $\pm$ 0.025  \\
13--16 & 1767 $\pm$ 51 & 851 $\pm$ 39 & 0.482 $\pm$ 0.027  \\
16--20 & 1269 $\pm$ 43 & 487 $\pm$ 30 & 0.384 $\pm$ 0.028  \\
20--25 & 642 $\pm$ 31 & 236 $\pm$ 22 & 0.368 $\pm$ 0.040   \\

\hline

\end{tabular}

\end{center}
\end{table}

\section{Systematic uncertainties} \label{sec:systematics}

Several types of systematic uncertainties are addressed. In particular,
we investigate possible  effects that could influence the measurement of
the numbers of $\Chione$ and $\Chitwo$ from data, the evaluation of
$\eoneetwo$ from the MC simulation, and the derivation of the
$R_\mathrm{p}$ ratio. In Table~\ref{table:sys} the various sources of systematic
uncertainties  and their contributions to the total
uncertainty are summarized. The following subsections describe how the various contributions are evaluated.

\begin{table*}[htbp]
\begin{center}

\topcaption{Relative systematic uncertainties on $R_\mathrm{p}$ for different ranges of \JPsi transverse momentum from different sources and the total uncertainty.}\label{table:sys}
\begin{tabular}{lcccccc}
\hline

$\pt(\JPsi)$ range [\GeVc]  & 7--9 & 9--11 & 11--13 & 13--16 & 16--20 & 20--25 \\
\hline

Source of uncertainty    &  \multicolumn{6}{c}{Systematic uncertainty (\%)}      \\
\hline

Background shape         & 1.4  & 1.5  & 0.9  & 1.2  & 1.8  & 2.4  \\
Signal shape             & 1.4  & 3.0  & 1.1  & 1.5  & 1.5  & 2.3  \\
Simulation sample size   & 2.6  & 2.0  & 2.2  & 2.4  & 3.1  & 4.8 \\
Choice of $\pt(\chic)$ spectrum   & 4.5  & 3.7  & 2.9  & 1.9  & 0.6  & 1.1  \\

\hline

Total uncertainty        & 5.5  & 5.4  & 3.9  & 3.6  & 4.0  & 5.9  \\

\hline
\end{tabular}

\end{center}
\end{table*}

\subsection{Uncertainty from the mass fit and \texorpdfstring{$\Chione$ and $\Chitwo$}{Chi[1] and Chi[2]}
  counting}

The measurement of the ratio $\theratio$ could be affected by the choice of the  functional form used for the maximum-likelihood fit. The use of an alternative  background parameterization, a fourth-order polynomial, results in systematically higher values of the ratio $\theratio$, while keeping the overall fit quality as high as in the default procedure. From the difference in the numbers of signal events using the two background parameterizations, we assign the systematic uncertainty  from the background modeling shown in Table~\ref{table:sys}.

We evaluate the systematic uncertainty related to  the parameterization of the signal shape by varying the parameters derived from the MC simulation within their uncertainties. The results  fluctuate within 1--3\% in the various transverse momentum ranges. We assign the systematic uncertainties from this source, as shown in Table~\ref{table:sys}.

The method to disentangle and count the $\Chione$ and $\Chitwo$ states is
validated by using  a \PYTHIA MC simulation
sample of inclusive \JPsi events, including those from $\chic$ decay, produced in pp collisions  and propagated
through the full simulation of the detector. The ratio $\theratio$ derived from the fit to the Q distribution of the reconstructed candidates in the simulation is consistent with the actual number of $\chic$ events contributing to the distribution, within the statistical uncertainty, for all \JPsi momentum ranges. Therefore, we do not assign any further systematic uncertainty in the determination of $\theratio$.

The stability of our analysis as a function of the number of primary vertices in the event has been investigated. The number of $\chic$ candidates per unit of integrated luminosity, once
trigger conditions are taken into account, is found to  be independent
of the instantaneous  luminosity, within the statistical uncertainties. In addition, the measured ratio $\theratio$  is found to be constant as a function of the number of primary vertices in the event, within the statistical uncertainties. Thus, no systematic uncertainty due to pileup is included in the final results.

\subsection{Uncertainty in the ratio of efficiencies}

The statistical uncertainty in the measurement of $\eoneetwo$ from the simulation, owing to the finite size of the MC sample, is taken as a systematic uncertainty, as shown in Table \ref{table:sys}.

Since the analysis relies on photon conversions, the
effect of a possible incorrect simulation of the tracker detector
material is estimated. Two modified material scenarios, i.e., special
detector geometries prepared for this  purpose, in which the total mass of the silicon tracker varies  by up to 5\% from the reference geometry,
are used to produce new  MC simulation samples~\cite{scenarios}.
With these models, local variations of the
radiation length  with respect to the reference simulation can be as large as $+8\%$ and $-3\%$. No significant difference in the ratio of efficiencies is observed and the corresponding systematic uncertainty is taken to be negligible.

Several choices of the generated $\pt(\chic)$ spectrum are investigated. In particular, the use of the measured \JPsi spectrum \cite{cms:jpsi} gives values that are compatible with the default $\psi(2S)$ spectrum used for the final result. The choice of the spectrum affects the values of $\eoneetwo$ only inasmuch as we perform an average measurement in each bin of $\pt(\JPsi)$, and the size of these bins is finite.
We choose to assign a conservative systematic uncertainty  by comparing the values of $\eoneetwo$ obtained with the  $\psi(2S)$ spectrum with those obtained in the case where the $\pt(\chic)$ spectrum is taken to be constant in each \pt bin. The corresponding systematic uncertainties are given in Table~\ref{table:sys}.

\subsection{\texorpdfstring{$\chi_c$}{Chi[c]} polarization}

The polarizations of the $\Chione$ and $\Chitwo$ are unknown.
Efficiencies are estimated under the assumption that the two states are unpolarized.
If the $\chic$ states are polarized, the resulting photon angular distribution and transverse momentum distributions will be affected. This can produce a change in the photon efficiency ratio $\eoneetwo$.

In order to investigate the impact of different polarization scenarios on the ratio of the efficiencies,
we reweight the unpolarized MC distributions to reproduce the theoretical $\chic$ angular distributions ~\cite{facciolimc,faccioli2} for different $\chic$ polarizations.  We measure the efficiency $\eoneetwo$ for the $\Chione$ being unpolarized or with helicity  $m_{\Chione} =0,\pm 1$, in combination with the $\Chitwo$ being unpolarized or having helicity $m_{\Chitwo} =0, \pm 2 $ in both the helicity and Collins--Soper~\cite{cs} frames. The ratio of efficiencies for the cases involving  $m_{\Chitwo} = \pm 1 $ is between  the cases with $m_{\Chitwo} = 0 $ and $m_{\Chitwo} = \pm 2 $.
 Tables~\ref{table:polarisationhx} and~\ref{table:polarisationcs} give the resulting $\eoneetwo$ values for each polarization scenario in different \JPsi transverse momentum bins for the two frames, relative to the value of the ratio for the unpolarized case. These tables, therefore, provide the correction that should be applied to the default value of $\eoneetwo$ in each polarization scenario and each  range of transverse momentum.

\begin{table*}[htbp]
\begin{center}
{
\topcaption{The efficiency ratio $\eoneetwo$ for different polarization
  scenarios in which the $\Chione$ is either unpolarized or has helicity $m_{\chi_{c1}}=0,\pm 1$ and the $\Chitwo$ is either unpolarized or has helicity $m_{\chi_{c2}}=0,\pm 2$ in the helicity frame, relative to the unpolarized case.}

\label{table:polarisationhx}
\begin{tabular}{ccccccc}
\hline
& \multicolumn{6}{c}{$\pt(\JPsi)[\GeVcns]$} \\
Polarization scenario ($m_{\chi_{c1}},m_{\chi_{c2}}$) & $7-9$ & $9-11$ & $11-13$ & $13-16$ & $16-20$ & $20-25$ \\
\hline
$(\mathrm{Unpolarized}, 0)$ & $0.89$ & $0.87$ & $0.85$ & $0.86$ & $0.85$ & $0.86$ \\
$(\mathrm{Unpolarized}, \pm 2)$ & $1.20$ & $1.20$ & $1.21$ & $1.20$ & $1.20$ & $1.17$ \\
$(0, \mathrm{Unpolarized})$ & $0.83$ & $0.84$ & $0.85$ & $0.85$ & $0.85$ & $0.86$ \\
$(\pm 1, \mathrm{Unpolarized})$ & $1.08$ & $1.07$ & $1.07$ & $1.07$ & $1.07$ & $1.07$ \\
$(0, 0)$ & $0.74$ & $0.73$ & $0.72$ & $0.73$ & $0.72$ & $0.74$ \\
$(0, \pm 2)$ & $1.00$ & $1.01$ & $1.03$ & $1.02$ & $1.02$ & $1.01$ \\
$(\pm 1, 0)$ & $0.95$ & $0.93$ & $0.91$ & $0.97$ & $0.90$ & $0.92$ \\
$(\pm 1, \pm 2)$ & $1.29$ & $1.29$ & $1.29$ & $1.28$ & $1.28$ & $1.25$ \\
\hline
\end{tabular}
}

\end{center}
\end{table*}

\begin{table*}[htbp]
\begin{center}
{
\topcaption{The values of $\eoneetwo$ for different polarization
  scenarios in the Collins--Soper frame, relative to the unpolarized case.}
\label{table:polarisationcs}
\begin{tabular}{ccccccc}

\hline
& \multicolumn{6}{c}{$\pt(\JPsi)[\GeVcns]$} \\
Polarization scenario ($m_{\chi_{c1}},m_{\chi_{c2}}$) & $7-9$ & $9-11$ & $11-13$ & $13-16$ & $16-20$ & $20-25$ \\
\hline
$(\mathrm{Unpolarized}, 0)$ & $1.04$ & $1.06$ & $1.08$ & $1.07$ & $1.08$ & $1.08$ \\
$(\mathrm{Unpolarized}, \pm 2)$ & $0.97$ & $0.95$ & $0.93$ & $0.93$ & $0.92$ & $0.92$ \\
$(0, \mathrm{Unpolarized})$ & $1.04$ & $1.05$ & $1.06$ & $1.07$ & $1.07$ & $1.06$ \\
$(\pm 1, \mathrm{Unpolarized})$ & $0.98$ & $0.97$ & $0.97$ & $0.96$ & $0.96$ & $0.97$ \\
$(0, 0)$ & $1.08$ & $1.12$ & $1.14$ & $1.15$ & $1.16$ & $1.14$ \\
$(0, \pm 2)$ & $1.01$ & $0.99$ & $0.98$ & $0.99$ & $0.98$ & $0.98$ \\
$(\pm 1, 0)$ & $1.02$ & $1.03$ & $1.04$ & $1.04$ & $1.04$ & $1.04$ \\
$(\pm 1, \pm 2)$ & $0.95$ & $0.92$ & $0.90$ & $0.90$ & $0.89$ & $0.89$ \\
\hline
\end{tabular}
}
\end{center}

\end{table*}

\subsection{Branching fractions}

The measurement of the prompt $\Chitwo$ to $\Chione$ production cross section ratio is affected by the uncertainties in the branching fractions of the two states into $\JPsi + \cPgg $. The quantity that is directly accessible in this analysis is $R_\mathrm{p}$, the product of the ratio  of the $\Chitwo$ to $\Chione$ cross sections and the ratio of the branching fractions.

In order to extract the ratio of the prompt production cross sections, we use the value of $1.76 \pm 0.10$ for ${\cal B}(\Chione \to \JPsi + \gamma) / {\cal B}(\Chitwo \to \JPsi + \gamma)$ as derived from the branching fractions and associated uncertainties reported in Ref.~\cite{PDG}.

\section{Results and discussion}

\begin{figure}[h!tbp]
    \centering
    \includegraphics[width=\cmsFigWidth]{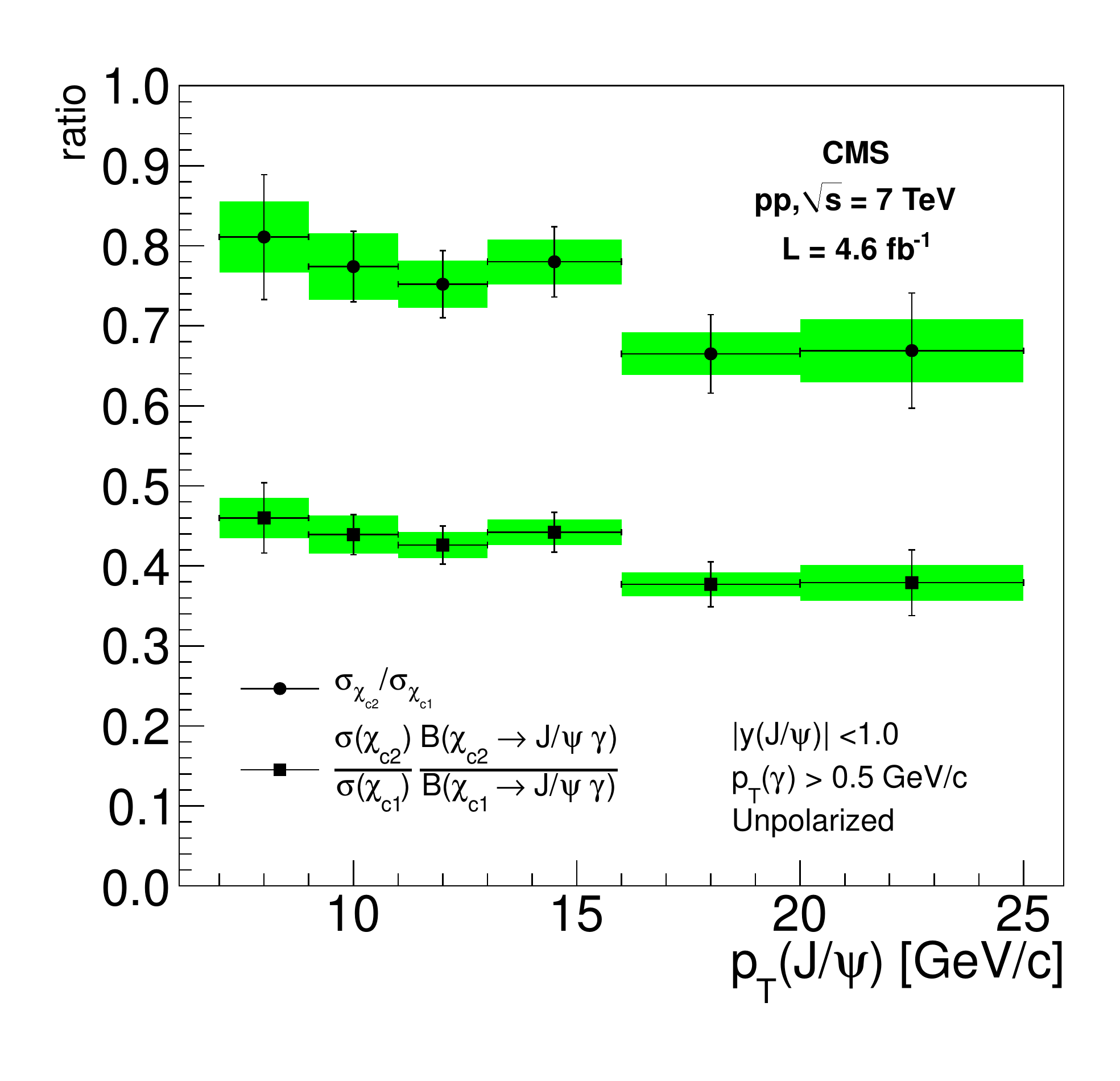}
    \caption{Ratio of the $\Chitwo$ to $\Chione$ production cross sections (circles) and ratio of the cross sections times the branching fractions to $\JPsi + \cPgg$ (squares) as a function of the \JPsi transverse momentum with the hypothesis of unpolarized production. The error bars correspond to the statistical uncertainties and  the green band corresponds to the systematic uncertainties. For the  cross section ratios,  the 5.6\% uncertainty  from the branching fractions is not included.}

    \label{fig:xsectionratio}
\end{figure}

The results of the measurement of the ratio $R_\text{p}$ and of the ratio of  the $\Chitwo$ to $\Chione$ prompt production cross sections for the kinematic range $\pt(\gamma) > 0.5$\GeVc and $|y(\JPsi)| < 1.0$ are
reported in Tables ~\ref{table:results2011AB:r} and \ref{table:results2011AB:s1s2}, respectively, for different ranges of $\pt(\JPsi)$.
The first uncertainty is statistical, the second  is systematic, and the third comes from the uncertainty in the branching fractions in the measurement of the cross section ratio. Separate columns are dedicated to the uncertainty derived from the extreme polarization scenarios in the helicity  and Collins--Soper frames, by choosing from Tables \ref{table:polarisationhx}  and \ref{table:polarisationcs} the scenarios that give  the largest variations relative to the unpolarized case. These correspond to $(m_{\Chione}, m_{\Chitwo}) = (\pm 1,\pm 2)$ and $(m_{\Chione}, m_{\Chitwo}) = (0,0)$ for both the helicity and Collins--Soper frames.
Figure~\ref{fig:xsectionratio} displays the results as a function of the \JPsi transverse momentum for the hypothesis of unpolarized production. The error bars represent the statistical uncertainties and the green bands the systematic uncertainties.

Our measurement of the ratio of the prompt $\Chitwo$ to $\Chione$ cross sections includes both directly produced  $\chic$ mesons and indirectly
produced ones from the decays of intermediate states.
To convert our result to the ratio of directly produced $\Chitwo$ to $\Chione$
mesons requires knowledge of the amount of feed-down from all possible
short-lived intermediate states that have a decay mode into $\Chitwo$ or
$\Chione$.  The largest known such feed-down contribution comes from the
$\psi(2S)$. Using the measured prompt \JPsi and $\psi(2S)$ cross sections in
pp collisions at 7\TeV \cite{cms:jpsi2}, the branching fractions for the decays
$\psi(2S) \to \Chionetwo + \gamma$ \cite{PDG}, and assuming the
same fractional $\chic$ contribution to the total prompt \JPsi production
cross section as measured in \Pp\Pap~collisions at 1.96\TeV
\cite{cdf:jpsitochic}, we estimate that roughly 5\%  of both our prompt
$\Chione$ and $\Chitwo$ sample comes from $\psi(2S)$ decays.  The correction in going from the prompt ratio to the direct ratio is about 1\%.  In comparing our
results with the theoretical predictions described below, we have not attempted to correct
for this effect since the uncertainties on the fractions are difficult to
estimate, the correction is much smaller than the statistical and
systematic uncertainties, and our conclusions on the comparisons with
the theoretical predictions would not be altered by a correction of this
magnitude.

\begin{figure}[htbp]
    \centering
    \includegraphics[width=0.49\textwidth]{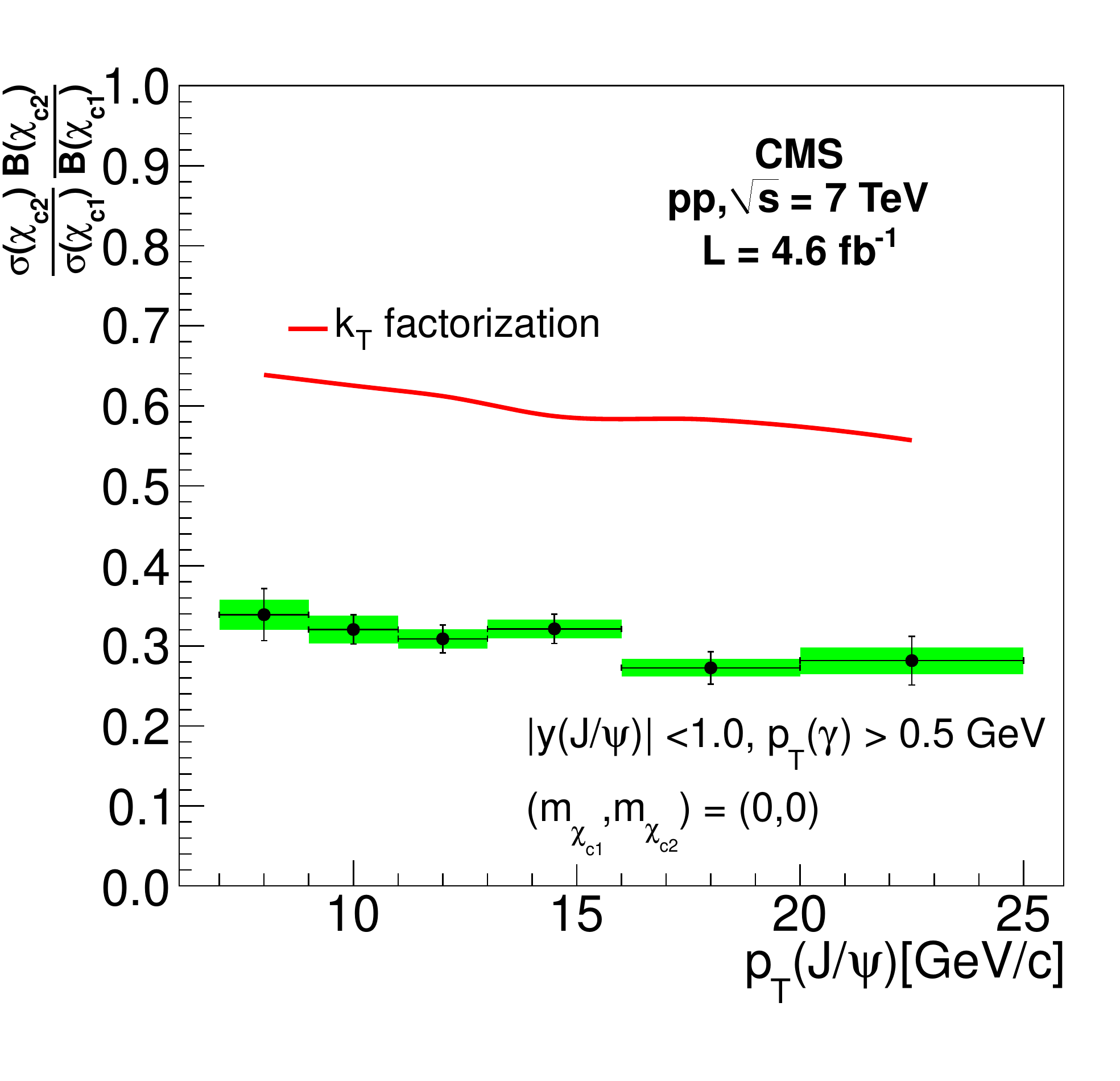}
    \includegraphics[width=0.49\textwidth]{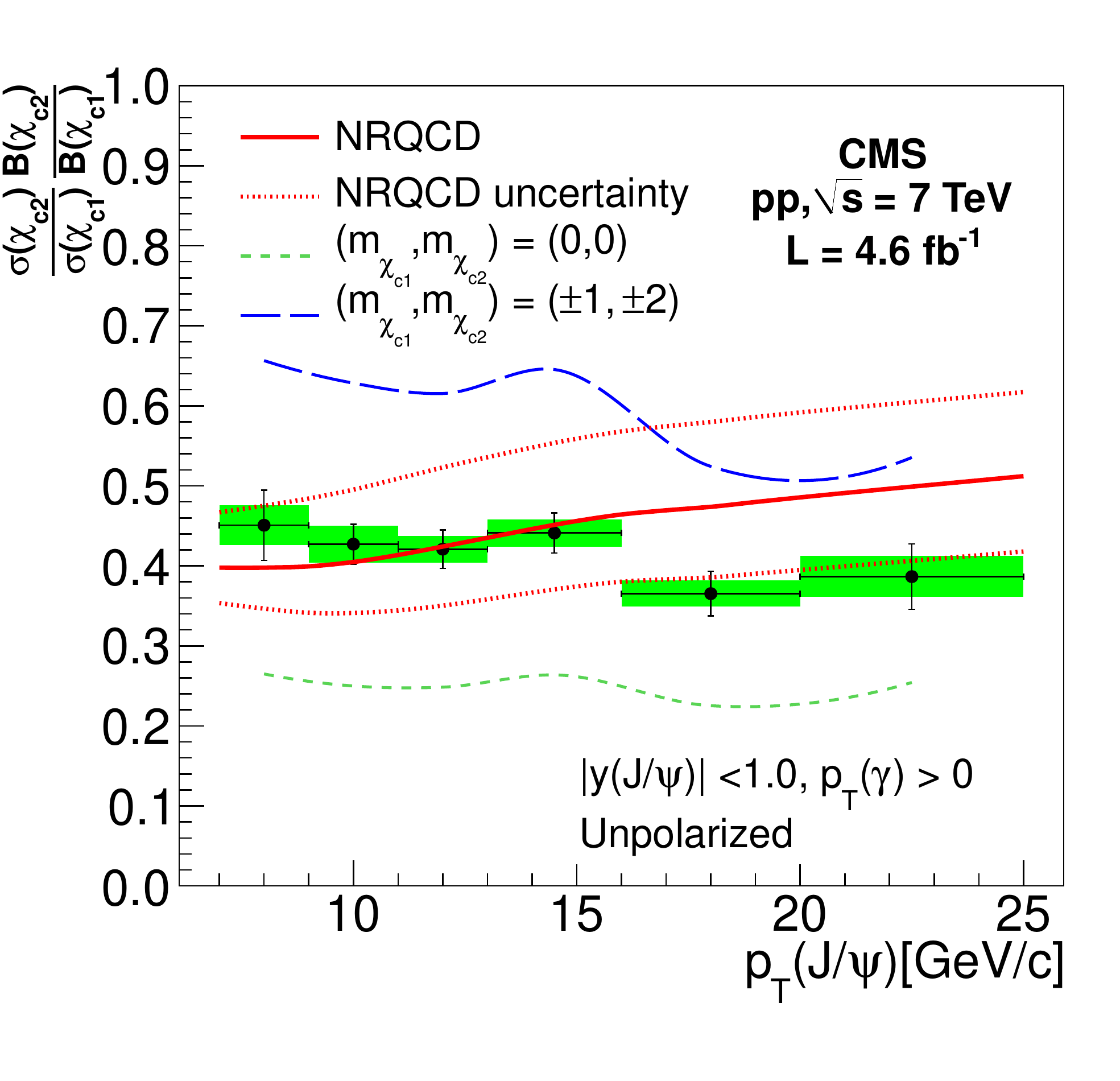}
    \caption{Comparison of the measured $ \frac{\sigma({\Chitwo}){\cal B}(\Chitwo) } { \sigma({\Chione}){\cal B}(\Chione)}$ values with theoretical predictions from the \kt-factorization \cite{baranov} (\cmsLeft) and NRQCD \cite{nlonrqcd} (\cmsRight) calculations (solid red lines). The error bars and green bands show the experimental statistical and systematic uncertainties, respectively. The measurements in the \cmsLeft plot use an acceptance correction assuming zero helicity for the $\chic$, as predicted by the \kt-factorization model. The measurements in the \cmsRight plot are corrected to match the kinematic range used in the NRQCD calculation and assume the $\chic$ are produced unpolarized. The measurements assuming two different extreme polarization scenarios are shown by the long-dashed blue and short-dashed green lines in the plot on the \cmsRight. The 1-standard-deviation uncertainties in the NRQCD prediction, originating from uncertainties in the color-octet matrix elements, are displayed as the dotted red lines.}
    \label{fig:theocomp}
\end{figure}

We compare our results with theoretical predictions derived from the \kt-factorization ~\cite{baranov} and NRQCD~\cite{nlonrqcd} calculations in Fig.~\ref{fig:theocomp}. The \kt-factorization approach predicts that both $\Chione$ and $\Chitwo$ are produced in an almost pure helicity-zero state in the helicity frame. Therefore, in our comparison, we apply the corresponding  correction on the ratio of efficiencies from Table \ref{table:polarisationhx}, amounting to a factor of  0.73, almost independent of $\pt$. The theoretical calculation is given in the same kinematic range ($\pt(\gamma) > 0.5$\GeVc, $|y(\JPsi)| < 1.0$) as our measurement.
There is no information  about the $\chic$ polarization from the NRQCD calculations, so we use the ratio of efficiencies  estimated in the unpolarized case for our comparison. The prediction is given in the kinematic range $\pt(\gamma) > 0$\GeVc, $|y(\JPsi)| < 1.0$. We use the same MC simulation described in Section \ref{sec:efficiency} to derive the small correction factor (ranging from  0.98 to 1.02 depending on \pt, with uncertainties from 1 to 4\%) needed to extrapolate the phase space of our measurement to the one used for the theoretical calculation. The uncertainty in the correction factor stemming from the assumption of the $\chic$ transverse momentum distribution is added as a systematic uncertainty. The values of $R_\text{p}$ after extrapolation are shown in Table \ref{table:results2011AB:rextrap}.
The comparison of our measurements with the \kt-factorization and NRQCD predictions  are  shown in the \cmsLeft and \cmsRight plots of Fig.~\ref{fig:theocomp}, respectively. The \kt-factorization prediction agrees well with the trend of $R_\text{p}$ versus transverse momentum of the \JPsi, but with a global normalization that is higher by about a factor two with respect to  our measurement. It is worth noting that this calculation assumes the same wave function for the $\Chione$ and the $\Chitwo$. On the other hand, the NRQCD prediction is compatible with our results within the experimental and theoretical uncertainties, though, since predictions for $\Chione$ or $\Chitwo$ polarizations were not provided, the level of agreement can vary considerably.

A direct comparison of our results with previous measurements, in particular from \cite{cdf:chi2chi1} and \cite{lhcb:chi}, is not straightforward, because of the different conditions under which they were carried out. Specifically, there are differences in the kinematical phase space considered and, in the case  of \cite{cdf:chi2chi1}, in the initial-state colliding beams and center-of-mass energy used. However, with these caveats, a direct comparison shows that the three results are compatible within their uncertainties. In particular, all three results confirm  the trend of a decreasing ratio of $\Chitwo$ to $\Chione$ production cross sections as a function of  $\pt(\JPsi)$, under the assumption that the $\Chitwo$ and  $\Chione$  polarizations  do not depend on $\pt(\JPsi)$.

\section{Summary}

Measurements have been presented  of the ratio
\[
R_\mathrm{p} \equiv
\frac{\sigma(\Pp\Pp \to \Chitwo +X )  \mathcal{B}(\Chitwo \to \JPsi + \gamma) }{ \sigma(\Pp\Pp \to \Chione +X )  \mathcal{B}(\Chione \to \JPsi + \gamma) }
\]
as a function of the \JPsi transverse momentum up to \cmsBreak $\pt(\JPsi) = 25\GeVc$ for the
kinematic range \ifthenelse{\boolean{cms@external}}{\linebreak[4]}{}$\pt(\gamma) > 0.5\GeVc$ and $|y(\JPsi)| < 1.0$
in $\Pp\Pp$ collisions at $\sqrt{s} = 7\TeV$ with a data sample corresponding to an integrated luminosity of  \usedlumi. The corresponding values for the ratio of the $\Chitwo$ to $\Chione$ production cross sections have been determined.

The results have also been shown after extrapolating the photon acceptance down to zero \pt.
The effect of several different $\chic$ polarization scenarios on the photon reconstruction efficiency has  been investigated  and taken into account in the comparison of the experimental results with two recent theoretical predictions.
This is among the most precise measurements of the $\chic$ production cross section ratio made in hadron collisions, and extends the explored \JPsi \pt range of previous results. These measurements will provide important input to and constraints on future theoretical calculations of quarkonium production, as recently discussed in \cite{likhoded} for the bottomonium family.

\begin{table*}[htbp]
\begin{center}
{
\topcaption{Measurements of  $ \frac { \sigma({\Chitwo})\mathcal{B}(\Chitwo)} { \sigma({\Chione{}})\mathcal{B}(\Chione)}$ for the given  $\pt(\JPsi)$ ranges in the fiducial  kinematic region $\pt(\gamma) > 0.5\GeVc $, $|y(\JPsi)| < 1.0$, assuming unpolarized $\chic$ production. The first uncertainty is statistical and the second is systematic. The last two columns report the additional  uncertainties  derived from  the extreme polarization scenarios in the helicity (HX)  and Collins--Soper (CS) frames.}
\label{table:results2011AB:r}
\renewcommand{\arraystretch}{1.4}
\begin{tabular}{lccc}
\hline
$\pt(\JPsi) [\GeVcns]$   &$ \frac { \sigma({\Chitwo})\mathcal{B}(\Chitwo)} { \sigma({\Chione{}})\mathcal{ B}(\Chione)}$  & HX & CS     \\

\hline
7--9  &  0.460$\pm$ 0.044 (stat.) $\pm$ 0.025 (syst.)  &$ ^{+0.136}_{-0.121}$&$^{+0.037}_{-0.023}$\\
9--11  &  0.439$\pm$ 0.025 (stat.) $\pm$ 0.024 (syst.) &$ ^{+0.128}_{-0.119}$&$^{+0.052}_{-0.035}$\\
11--13  &  0.426$\pm$ 0.024 (stat.) $\pm$ 0.017 (syst.)&$ ^{+0.125}_{-0.117}$&$^{+0.059}_{-0.042}$\\
13--16  &  0.442$\pm$ 0.025 (stat.) $\pm$ 0.016 (syst.)&$ ^{+0.125}_{-0.121}$&$^{+0.065}_{-0.044}$\\
16--20  &  0.377$\pm$ 0.028 (stat.) $\pm$ 0.015 (syst.)&$ ^{+0.106}_{-0.104}$&$^{+0.059}_{-0.042}$\\
20--25  &  0.379$\pm$ 0.041 (stat.) $\pm$ 0.022 (syst.)&$ ^{+0.094}_{-0.097}$&$^{+0.055}_{-0.040}$\\
\hline

\end{tabular}
}

\end{center}
\end{table*}

\begin{table*}[htbp]
\begin{center}
{
\topcaption{Measurements of $\sigma({\Chitwo}) / \sigma({\Chione})$ for the given  $\pt(\JPsi)$ ranges  derived using the branching fractions from Ref.~\cite{PDG}, assuming unpolarized $\chic$ production. The first uncertainty is statistical, the second is systematic,
 and the third from  the branching fraction uncertainties. The last two columns report the uncertainties  derived from  the extreme polarization scenarios in the helicity (HX) and Collins--Soper (CS) frames.}

\label{table:results2011AB:s1s2}
\renewcommand{\arraystretch}{1.4}
\begin{tabular}{lccc}
\hline
$\pt(\JPsi) [\GeVcns]$  & $\sigma({\Chitwo})/\sigma({\Chione})$ & HX & CS     \\
\hline
7--9 &  0.811$\pm$0.078 (stat.) $\pm$ 0.045 (syst.)  $\pm$ 0.046(BR)  &$ ^{+0.239}_{-0.213}$&$^{+0.066}_{-0.041}$  \\
9--11 &  0.774$\pm$0.044 (stat.) $\pm$ 0.042 (syst.)  $\pm$ 0.044(BR) &$ ^{+0.225}_{-0.209}$&$^{+0.092}_{-0.061}$ \\
11--13 &  0.752$\pm$0.042 (stat.) $\pm$ 0.029 (syst.)  $\pm$ 0.043(BR) &$ ^{+0.221}_{-0.207}$&$^{+0.105}_{-0.074}$\\
13--16 &  0.78$\pm$0.044 (stat.) $\pm$ 0.028 (syst.)  $\pm$ 0.044(BR) &$ ^{+0.221}_{-0.213}$&$^{+0.115}_{-0.078}$ \\
16--20 &  0.665$\pm$0.049 (stat.) $\pm$ 0.027 (syst.)  $\pm$ 0.038(BR)&$ ^{+0.187}_{-0.184}$&$^{+0.104}_{-0.074}$\\
20--25 &  0.669$\pm$0.072 (stat.) $\pm$ 0.039 (syst.)  $\pm$ 0.038(BR)&$ ^{+0.165}_{-0.172}$&$^{+0.096}_{-0.070}$\\

\hline
\end{tabular}
}

\end{center}
\end{table*}

\begin{table*}[htbp]
\begin{center}
{
\topcaption{Measurements of  $ \frac { \sigma({\Chitwo})\mathcal{B}(\Chitwo)} { \sigma({\Chione{}})\mathcal{ B}(\Chione)}$ for the given  $\pt(\JPsi)$ ranges after extrapolating  the measurement  to the kinematic region $\pt(\gamma) > 0$ and assuming unpolarized $\chic$ production. The first uncertainty is statistical and  the second is systematic. The last column reports the largest variations due changes in the assumed $\chic$ polarizations. }
\label{table:results2011AB:rextrap}
\renewcommand{\arraystretch}{1.4}
\begin{tabular}{lcc}
\hline
$\pt(\JPsi) [\GeVcns]$   &$ \frac { \sigma({\Chitwo})\mathcal{B}(\Chitwo)} { \sigma({\Chione{}})\mathcal{ B}(\Chione)}$  & Polarization  \\

\hline
7--9    &  0.451  $\pm$  0.043  (stat.) $\pm$  0.025  (syst.) &  $^{+ 0.137} _{- 0.153}$\\
9--11   &  0.427  $\pm$  0.024  (stat.) $\pm$  0.023  (syst.) &  $^{+ 0.134} _{- 0.144}$\\
11--13  &  0.421  $\pm$  0.024  (stat.) $\pm$  0.017  (syst.) &  $^{+ 0.133} _{- 0.142}$\\
13--16  &  0.441  $\pm$  0.025  (stat.) $\pm$  0.017  (syst.) &  $^{+ 0.138} _{- 0.143}$\\
16--20  &  0.365  $\pm$  0.027  (stat.) $\pm$  0.016  (syst.) &  $^{+ 0.114} _{- 0.115}$\\
20--25  &  0.387  $\pm$  0.042  (stat.) $\pm$  0.026  (syst.) &  $^{+ 0.109} _{- 0.105}$\\

\hline
\end{tabular}
}

\end{center}
\end{table*}

\section*{Acknowledgments}

{\tolerance=800
\hyphenation{Bundes-ministerium Forschungs-gemeinschaft
  Forschungs-zentren} The authors would like to thank Sergey Baranov for providing theoretical calculations in the \kt-factorization scheme and Kuang-Ta Chao and Yan-Qing Ma for their NRQCD predictions.

We wish to congratulate our colleagues in the
CERN accelerator departments for the excellent performance of the LHC
machine. We thank the technical and administrative staff at CERN and
other CMS institutes. This work was supported by the Austrian Federal
Ministry of Science and Research; the Belgium Fonds de la Recherche
Scientifique, and Fonds voor Wetenschappelijk Onderzoek; the Brazilian
Funding Agencies (CNPq, CAPES, FAPERJ, and FAPESP); the Bulgarian
Ministry of Education and Science; CERN; the Chinese Academy of
Sciences, Ministry of Science and Technology, and National Natural
Science Foundation of China; the Colombian Funding Agency
(COLCIENCIAS); the Croatian Ministry of Science, Education and Sport;
the Research Promotion Foundation, Cyprus; the Estonian Academy of
Sciences and NICPB; the Academy of Finland, Finnish Ministry of
Education and Culture, and Helsinki Institute of Physics; the Institut
National de Physique Nucl\'eaire et de Physique des Particules~/~CNRS,
and Commissariat \`a l'\'Energie Atomique et aux \'Energies
Alternatives~/~CEA, France; the Bundesministerium f\"ur Bildung und
Forschung, Deutsche Forschungsgemeinschaft, and Helmholtz-Gemeinschaft
Deutscher Forschungszentren, Germany; the General Secretariat for
Research and Technology, Greece; the National Scientific Research
Foundation, and National Office for Research and Technology, Hungary;
the Department of Atomic Energy and the Department of Science and
Technology, India; the Institute for Studies in Theoretical Physics
and Mathematics, Iran; the Science Foundation, Ireland; the Istituto
Nazionale di Fisica Nucleare, Italy; the Korean Ministry of Education,
Science and Technology and the World Class University program of NRF,
Korea; the Lithuanian Academy of Sciences; the Mexican Funding
Agencies (CINVESTAV, CONACYT, SEP, and UASLP-FAI); the Ministry of
Science and Innovation, New Zealand; the Pakistan Atomic Energy
Commission; the Ministry of Science and Higher Education and the
National Science Centre, Poland; the Funda\c{c}\~ao para a Ci\^encia e
a Tecnologia, Portugal; JINR (Armenia, Belarus, Georgia, Ukraine,
Uzbekistan); the Ministry of Education and Science of the Russian
Federation, the Federal Agency of Atomic Energy of the Russian
Federation, Russian Academy of Sciences, and the Russian Foundation
for Basic Research; the Ministry of Science and Technological
Development of Serbia; the Ministerio de Ciencia e Innovaci\'on, and
Programa Consolider-Ingenio 2010, Spain; the Swiss Funding Agencies
(ETH Board, ETH Zurich, PSI, SNF, UniZH, Canton Zurich, and SER); the
National Science Council, Taipei; the Scientific and Technical
Research Council of Turkey, and Turkish Atomic Energy Authority; the
Science and Technology Facilities Council, UK; the US Department of
Energy, and the US National Science Foundation.

Individuals have received support from the Marie-Curie programme and
the European Research Council (European Union); the Leventis
Foundation; the A. P. Sloan Foundation; the Alexander von Humboldt
Foundation; the Belgian Federal Science Policy Office; the Fonds pour
la Formation \`a la Recherche dans l'Industrie et dans l'Agriculture
(FRIA-Belgium); the Agentschap voor Innovatie door Wetenschap en
Technologie (IWT-Belgium); the Council of Science and Industrial
Research, India; and the HOMING PLUS programme of Foundation for
Polish Science, cofinanced from European Union, Regional Development
Fund.\par}

\bibliography{auto_generated}   

\cleardoublepage \appendix\section{The CMS Collaboration \label{app:collab}}\begin{sloppypar}\hyphenpenalty=5000\widowpenalty=500\clubpenalty=5000\textbf{Yerevan Physics Institute,  Yerevan,  Armenia}\\*[0pt]
S.~Chatrchyan, V.~Khachatryan, A.M.~Sirunyan, A.~Tumasyan
\vskip\cmsinstskip
\textbf{Institut f\"{u}r Hochenergiephysik der OeAW,  Wien,  Austria}\\*[0pt]
W.~Adam, E.~Aguilo, T.~Bergauer, M.~Dragicevic, J.~Er\"{o}, C.~Fabjan\cmsAuthorMark{1}, M.~Friedl, R.~Fr\"{u}hwirth\cmsAuthorMark{1}, V.M.~Ghete, J.~Hammer, N.~H\"{o}rmann, J.~Hrubec, M.~Jeitler\cmsAuthorMark{1}, W.~Kiesenhofer, V.~Kn\"{u}nz, M.~Krammer\cmsAuthorMark{1}, I.~Kr\"{a}tschmer, D.~Liko, I.~Mikulec, M.~Pernicka$^{\textrm{\dag}}$, B.~Rahbaran, C.~Rohringer, H.~Rohringer, R.~Sch\"{o}fbeck, J.~Strauss, A.~Taurok, W.~Waltenberger, G.~Walzel, E.~Widl, C.-E.~Wulz\cmsAuthorMark{1}
% ...
\vskip\cmsinstskip
\dag:~Deceased\\
1:~~Also at Vienna University of Technology, Vienna, Austria\\
% ...

\end{sloppypar}
\end{document}